\documentclass[11pt,fleqn,leqno]{article}
\author{Mihaela D. Iftime}
\title{Space-time distributions}
\date{31~Aug~98}

\begin{document}
\maketitle

\begin{abstract}
The space-time foliation $\Sigma$ compatible with the gravitational field $g$ on a 4-manifold $M$ determines a fibration $\pi$ of $M$,
$\pi : M\longrightarrow \mathcal{N}$ is a surjective submersion over the 1-dimensional leaves space $\mathcal{N}$.
$M$ is then written as a disjoint union of the leaves of $\Sigma$, which are 3-dimensional spacelike surfaces on $M$.

The decomposition, $TM=\Sigma\oplus T^{0} M$, also implies that we can define a lift of the curves on $\mathcal{N}$ to curves (non-spacelike)
on $M$.

The stable causality condition $M$ \cite{hw:blackhole} coincides with $\Sigma$ being a causal space-time distribution, generated by
an exact timelike 1-form $\omega^{0}=\mathrm{d}t$ where $t$ is some real function on $M$. In this case $M$ is written as a disjoint
union of a family of spacelike 3-surfaces of constant $t$, which cover $D^{+}(S)$ of a initial 3-surface $S$ of $M$.
\end{abstract}

\vspace{6pt}
{\bf AMS Classification numbers:}

Primary: 83C99

Secondary: 53B30 53C50 53C12 57R30

\vspace{6pt}

\pagebreak

\section {Introduction}\label{sec:into}
{\bf Note:} We will denote by small Latin indices $a, b, \ldots$ numbers ranging from 0 to 3 and the small Greek indices $\alpha, \beta, \ldots$ numbers
ranging from 1 to 3.\vspace{6pt}

The space-time structure on a 4-dimentional manifold $M$ is a uniquely defined by a triplet
$(g^R , g , \Sigma )$, where $g$ and $g^R$ represent a Riemannian, respectively a Lorentzian metric on $ M$ and 
$\Sigma \subset \tau_{M}$, the space-time distribution on $M$ compatible with the gravitational field $g$,
is a 3-dimentional spatial tangent subbundle on $M$ orthocomplement (with respect both $g$ and $g^{R}$) of the 1-dimentional timelike subbundle $T^{0} M$ generated by the timelike globally defined vector field  
$h^{0}\in \mathcal{X}(M)$,  
$ g^{R}( h^{0}\, ,h^{0})=-g( h^{0}\, ,h^{0})= 1 $
which indicates the time orientation (locally)
at every point of $M$.

If $ g^{R}$ is a Riemannian metric on $M$ (which exists on every paracompact space) and a time-orientation defined by $ h^{0}\in \mathcal{X}(M) $ then we have a Lorentz metric on $M$ determined by the expression :
\begin{equation}
g(X, \, Y)=g^{R}(X, \, Y)-2g^{R}(X, \, h^{0})g^{R}(Y, \,h^{0})
\quad \forall  X,\, Y \in \mathcal{X}(M)
\end{equation}
Locally, relating  to an orthonormal frame 
$ \{e_{a}\}=\{ h^{\alpha }, \, h^{0}\} $ 
we find:
\begin{equation}
g_{ab}=g_{ab}^{R}-2\omega _{a}^{0}\omega_{b}^{0}, \label{eq:omega}
\end{equation}
where $\omega^{0}=g^{R}(h^{0}, \, \cdot )$ is the unitary timelike 1-form.

The Pfaff equation $\omega^{0}=0$ generates the 3-dimensional spatial spaces $\Sigma_{p}$  in each fibre $T_{p} M$ of the tangent bundle $\tau_{M}$ and these subspaces are orthogonal to the timelike field $h^{0}$ in $p\in M$.

The tensor 
\begin{equation}
 h_{ab} = g_{ab} + \omega_{a}^{0}\omega_{b}^{0} \label{eq:h}
\end{equation}
defines a Riemannian metric on the tangent subbundle $\Sigma$.

We have the following decomposition of tangent bundle:
\begin{equation}
 TM = \Sigma\oplus T^{0} M \label{eq:desc}
\end{equation}
So, given a Riemannian metric $g_{ab}^{R}$ on $M$, each 3-dimensional distribution $\Sigma$ on $M$ generated
by a nonzero 1-form $\sigma$ on $M$ represents a space-time distribution compatible with a gravitational field $g$
defined by the expression (\ref{eq:omega}), where we take $\omega^{0} = -\frac{\sigma}{g^{R}(\sigma,\, \sigma)}$.

Inverse, given a gravitational field $g$ on $M$ we find a spacetime distribution $\Sigma$ compatible with $g$ and
a Riemannian metric $g^{R}$ which satisfy the expression (\ref{eq:omega}).

Two such different triplets $(g_{i},\, g_{i}^{R},\, \Sigma_{i}), \enspace i=1,\, 2$ will describe the same space-time
structure on $M$ if the space-time distributions coincide, $\Sigma_{1} = \Sigma_{2}$.

The geometric equivalence principle --- which states that we may introduce an orthonormal tetrad field
$\{h^{\alpha},\, h^{0}\}$ at each point of $M$ --- is also sufficient for defining the space-time structure on a
4-manifold $M$.

The metric $g_{ab}$ can be transformed to
\begin{equation}
 g_{ab} = \eta_{ab} = diag(1,\, 1,\, 1,\, -1)
\end{equation}
at any point of $M$.

Also for any point $p\in M$ and its normal convex neighbourhood $U_{p} \subset M, \enspace
U_{p} = \exp_{p}W_{0}$, where $W_{0}\subset T_{p} M$ is an open neighbourhood of the origin of $T_{p}M$, we can have
a causal structure; all events $q$ in $U_{p}$ are divided into three groups according whether the causal relation to $p$ is
timelike, null or spacelike and these relations are independent of the choice of coordinate system. But we don't know how to
split an arbitrary space-time in space and time globally.

An integrable distribution $$\Sigma = \bigcup_{p\in M} \Sigma_{p}$$ is a spacetime foliation of spatial hypersurfaces
$S$ of $M; \enspace \Sigma_{p} = T_{p} S, \enspace \forall p\in S$.

In this case, at every point $p\in M$ we can define a local coordinate neighbourhood $U_{i}\subset M$ with local
adapted coordinates $(x^{a})$ so that the spacelike vector fields $\{\frac{\partial}{\partial x^{\alpha}}\}$ generate
$\Sigma$ locally.

The Pfaff equation $x^{0} = \mathrm{const.}$ locally defines the spatial hypersurfaces $S$ of $M$ (i.e. the maximal integral manifolds
or the leaves of $\Sigma$).

We have, $$\mathcal{U} = \{ U_{i},\enspace t_{i}:\tau_{M}^{-1}(U_{i})\longrightarrow U_{i}\times \mathrm{R}^{4}\}$$ a bundle
atlas on $TM$ which is compatible with the subbundle $\Sigma\subset \tau_{M}$, so that
$$\mathcal{U}^{\Sigma} = \{ U_{i},\enspace t_{i}^{-1}:\tau_{M}^{-1}(U_{i}) \cap \Sigma \longrightarrow U_{i}\times \mathrm{R}^{3}\}$$
is a bundle atlas on $\Sigma$, $t_{i}(\tau_{M}^{-1}(U_{i}) \cap \Sigma)=U_{i} \times \mathrm{R}^{3}$.

We can have directly: for any adapted local coordinate system $(U_{i},\, (x^{a}))$, then $(x^{a},\, \dot x^{a})$ are local bundle
coordinates on $TM$ and $\{\frac{\partial}{\partial x^{a}}\}$ are the dual sections. Then, from the local trivialisation of $\tau_{M}$
\begin{equation}
 t_{i} = ( \tau_{{M}_{\bigr|\tau_{M}^{-1}(U_{i})}},\, \dot x^{a}):\tau_{M}^{-1}(U_{i}) \longrightarrow U_{i} \times \mathrm{R}^{4}
\end{equation}
we obtain a local trivialisation on subbundle $\Sigma$
\begin{equation}
 t_{i}^{\Sigma} = ( \tau_{{M}_{\bigr|\tau_{M}^{-1}(U_{i})\cap \Sigma}},\, \dot x^{\alpha}): \tau_{M}^{-1}(U_{i})\cap \Sigma 
\longrightarrow U_{i} \times \mathrm{R}^{3}.
\end{equation}

\section{Family of spatial hypersurfaces of spacetime}
A space-time foliation $\Sigma $ on $M$ generates a fibration $\pi$ as follows: we define an equivalence relation on $M$,
$ x, y \in M$, $x\sim y$ if $x, \, y$ are in the same leaf of the foliation.

We denote by $\mathcal {N}=M_{\vert \sim }$ the space of leaves which has the natural quotient structure  of 1-dimensional differentiable manifold, so that $\pi:  M \longrightarrow \mathcal{N} $ is a surjective submersion. The leaves of the foliation $\Sigma$ 
coincide with the fibres of $\pi$, $\pi ^{-1}(\nu)=S_{\nu}$ and $\Sigma=\mathrm{Ker} \pi _{\ast}$ . We also may suppose that
the fibres of $\pi$ are simply connected even if it is not a necessary condition.

We can define an vertical lift of the tangent vectors on $\mathcal{N}$ to tangent vectors on $M$ which belong to $\Sigma$.

For  $ \bar {X} \in T_{\nu}\mathcal{N} $ its vertical lift is $X^{v} \in T_{p} M, \enspace {X^{v}}^a=h_{b}^{a}(p)  X^{b}, \enspace
\forall X \in T_{p}M$, where $\pi_{\ast}X=\bar X$.

Here
\begin{eqnarray}
 h = (h_{b}^{a}):TM \longrightarrow TM\\
 h_{b}^{a} = g^{ac} h_{cb} \nonumber
\end{eqnarray}
is an endomorphism on $\tau_{M}$, a projection operator, i.e. $h_{b}^{a} h_{c}^{b}=h_{c}^{a}$, which projects any vector
$X \in T_{p}M$ into its part lying in the subspace $\Sigma_{p}$,
\begin{equation}
 X^{a} = h_{b}^{a} X^{b} - {h^{0}}^{a} \omega_{b}^{0} X^{b}
\end{equation}
and any $X \in \mathcal{X}(\mathcal{N})$ then $ X^{v}\in \mathcal{V}(\Sigma ) $ defined by $X^{v}=h(X)$ for any
 $X\in \mathcal{X}(M)$ so that $\pi _{\ast }X=\bar X $.

Here $ \mathcal{V}(\Sigma) =\{ X\in \mathcal{X}_{loc}(M)\enspace/\, X_{p}\in \Sigma _{p},\, \forall p  \}$.

The Lorentzian (metric) connection $\nabla$ on $ M$ induces by restriction a connection $\nabla ^{S}$ on each leaf of   
$ \Sigma $ ,
\begin{equation}
\nabla_{X^{v}}Y^{v}=\nabla _{X^{v}}^{S} h(Y) = h(\nabla _{X^{v}}Y) \in \Sigma .
\end{equation}
We used
\begin{equation}
0=( \nabla _{X^{v}}h )Y=\nabla _{X^{v}}h(Y)-h(\nabla _{X^{v}}Y)
\end{equation}
and
\begin{equation}
h_{ab;c}=(g_{ab}+\omega^{0}_{a}\omega^{0}_{b})_{;c}=0
\end{equation}

It is worth notice that the space-time is a disjoint union of the leaves of the foliation $\Sigma$, i.e.
$$ M=\bigcup _{\nu \in \mathcal{N}}S_{\nu}$$
where $ S_{\nu}=\pi^{-1}(\nu)$ represents a family of spatial hypersurfaces of $M$.
From decomposition (\ref {eq:desc})  we can define a transport of the those fibres $ S_{\nu} $ along the curves $ \gamma (\tau ) $
on $\mathcal{N}$. That is to say, a curve 
$\gamma (\tau) :\lbrack 0,\,1  \rbrack\longrightarrow \mathcal{N}, \enspace \gamma(0)=\nu_{0}, \, \gamma(1)=\nu_{1}$ could be lifted on $M$ from each point  of the fibre $x_{i} \in \pi^{-1}(\nu_{0}) = S$ to a nonspatial future directed curve on $M$ of end-point
$y_{i} \in \pi^{-1}(\nu_{1}) =S_{\nu_{1}}$.

This way we obtain an 1:1 correspondence between $S$ and $S_{\nu_{1}}$. Then we have a family of maps (homeomorphisms)
\begin{equation}
 \alpha_{\gamma}: S_{\gamma(0)} \longrightarrow S_{\gamma(1)}, \quad \alpha_{\gamma}:x_{i} \longmapsto y_{i}
\end{equation}
satisfying natural conditions:
$ \alpha_{\gamma_{1} \gamma_{2}} = \alpha_{\gamma_{1}} \alpha_{\gamma_{2}}$, \enspace
$ \alpha_{\gamma^{-1}} = (\alpha_{\gamma})^{-1} $, \enspace
$ \alpha_{\gamma} = \mathrm{id}$ for $\gamma = \mathrm{const.} $, \enspace
$ \alpha_{\gamma}$ is independent of the choice of the parameter $\tau$ on $\gamma$.

On the other hand $\mathcal{N}$ is 1-dimensional (connected) manifold. This implies that $\mathcal{N}$ must be diffeomorphic
with an interval $\lbrack0,\,1\rbrack$ or with circle $S^{1}$.

Supposing $\mathcal{N}$ has a fixed orientation given by the unique unitary $\bar X \in \mathcal{X}(\mathcal{N})$, then
$\mathcal{N} = \gamma_{\bar X}(\lbrack0,\,1\rbrack)$, where $\gamma_{\bar X}:\lbrack0,\,1\rbrack \longrightarrow \mathcal{N}$ is
the integral curve of $\bar X$. We will have the two cases: when $\gamma_{\bar X}$ is an injective integral curve,
$\mathcal{N} = \lbrack0,\,1\rbrack$ and when $\gamma_{\bar X}$ admits cross-points (i. e. $\gamma_{\bar X}$ may intersect
itself), $\mathcal{N} = S^{1}$.

Lifting $\gamma_{\bar X}$ from each point $x_{i} \in \pi^{-1}(\nu_{0}) = S$ we obtain a family of timelike future directed curves 
$\tilde{\gamma}_{\bar X} : \lbrack0,\,1\rbrack \longrightarrow M$, $\dot{\tilde{\gamma}}_{\bar X}(\tau) \in T_{p}^{0} M, \enspace \pi(p) = \nu$ which start on $S$ and intersect each leaf $S_{\nu}$ orthogonally.

The stable causality condition holds on $M$ \footnote{This condition of stability is equivalent to the existence of a function $f$ on $M$,
whose $\nabla f$ is everywhere timelike.} means that the space-time foliation $\Sigma$ is causal, being generated by a globally defined
timelike 1-form $\omega^{0} \in \wedge^{1} M$ which is exact, i.e. $\omega^{0} = \mathrm{d}t$ where $t$ is some real function on $M$.

In this case, the leaves of $\Sigma$ are spatial hypersurfaces of $M$ of constant $t$, and $$M = \bigcup_{\nu \in \mathcal{N}} S_{\nu}$$ is the disjoint union of a family of spatial hypersurfaces $S_{\nu}$ of constant $t$ on $M$.

If we suppose that the initial surface $S$ is a partial Cauchy surface of $M$ we obtain $M$ as a disjoint union of such homeomorphic
partial Cauchy surfaces $S_{\nu}$, which in addition satisfy the following conditions: for $\nu_{i} < \nu_{j}$ then $S_{\nu_{j}}$ lies
to the future of $S_{\nu_{i}}$ and they cover $D^{+}(S)$.

In the case when $\mathcal{N}$ is diffeomorphic with $S^{1}$, provided that $S$ is a partial Cauchy surface on $M$, a curve
$\gamma_{\bar X}:\lbrack0,\,1\rbrack \longrightarrow \mathcal{N}, \enspace \gamma_{\bar X}(0) = \gamma_{\bar X}(1) \in S^{1}$ is
lifted from a point $x_{i}$ of $S$ to a closed timelike curve $\tilde{\gamma}$ on $M$ with $\tilde{\gamma}(0) = x_{i} = \tilde{\gamma}(1)$.

\section{Some discussion}

Physically, it seems more convenient to choose the surfaces $S_{\nu}$ so that they intersect $\mathcal{I}^{+}$
\cite{hw:blackhole}.
This means that every $S_{\nu}$ tend asymptotically to null surfaces.

Also by assumption of strong asymptotic predictability (i.e. there exists a partial Cauchy surface $S$ of $M$ such that (a)
$\mathcal{I}^{+}$ lies in the boundary of $D^{+}(S)$ and (b) $J^{+}(S) \cap \partial J^{-}(\mathcal{I}^{+})$ lies in
$D^{+}(S)$ ) for some $\nu$, the surfaces $S_{\nu_{i}}$ for $\nu_{i} > \nu$ will intersect the event horizon
$\partial J^{-}(\mathcal{I}^{+})$. Then a connected component of the nonempty set
$S_{\nu_{i}}\backslash J^{-}(\mathcal{I}^{+})$ will represent a black hole on the surface $S_{\nu_{i}}$.


\begin{thebibliography}{9}
\bibitem{bg:foliation} {\bf Berger, B.K., Chru'sciel, P.T., Isenberg, J., Moncrief, V.},
 \emph{Global foliations of vacuum spacetimes with T2 isometry}, Ann. Phys. (N. Y.), 260, 117-148, (1997)
\bibitem{hw:blackhole} {\bf Hawking, S.W.},
 \emph{Hawking on the Big Bang and Black Holes}, World Scientific (1993)
\bibitem{iri:foliation} {\bf Iriondo, M., Leguizam'on, E., and Reula, O.},
 \emph{The Newtonian Limit on Asymptotically Null Foliations}, Online Los Alamos Archive Preprint (1997)
\bibitem{rd:foliation} {\bf Rendall, A.D.},
 \emph{Existence and non-existence results for global constant mean curvature foliations}, Online Los Alamos Archive Preprint, (1996)
\bibitem{sd:gauge} {\bf Sardanashvily, G., Zakharov, O.},
 \emph{Gauge gravitation theory}, World Scientific (1992)
\end{thebibliography}
\end{document}